\documentclass[pra,twocolumn,superscriptaddress,10pt,floatfix]{revtex4}
\usepackage[pdftex,plainpages=false,colorlinks=true,citecolor=blue,linkcolor=blue,urlcolor=blue,filecolor=green,bookmarksopen=true]{hyperref}
\usepackage[utf8]{inputenc}
\usepackage{amsmath}
\usepackage[caption = false]{subfig}
\usepackage{graphicx,epstopdf}
\usepackage{blindtext}
\usepackage{lipsum}
\usepackage{amsfonts}
\usepackage{bbm}
\usepackage{amssymb}
\usepackage{enumerate}
\usepackage{color}
\usepackage{latexsym}
\usepackage{times,txfonts}

\newcommand{\Ical}{\mathcal{I}}

\newcommand{\Fcal}{\mathcal{F}}

\newcommand{\ket}[1]{| #1 \rangle}

\newcommand{\interpro}[2]{\langle #1 | #2 \rangle}
\newcommand{\bra}[1]{\langle #1 |}

\begin{document}

\title{Experimental Implementation of Generalized Transitionless Quantum Driving}

\author{Chang-Kang Hu}
\thanks{These two authors contributed equally to this work}
\affiliation{CAS Key Laboratory of Quantum Information, University of Science and Technology of China, Hefei, 230026, People’s Republic of China}
\affiliation{Synergetic Innovation Center of Quantum Information and Quantum Physics,University of Science and Technology of China, Hefei, 230026, People’s Republic of China}

\author{Jin-Ming Cui}
\thanks{These two authors contributed equally to this work}
\affiliation{CAS Key Laboratory of Quantum Information, University of Science and Technology of China, Hefei, 230026, People’s Republic of China}
\affiliation{Synergetic Innovation Center of Quantum Information and Quantum Physics,University of Science and Technology of China, Hefei, 230026, People’s Republic of China}

\author{Alan C. Santos}
\email{ac\_santos@id.uff.br}
\affiliation{Instituto de F\'{i}sica, Universidade Federal Fluminense, Av. Gal. Milton Tavares de Souza s/n, Gragoat\'{a}, 24210-346 Niter\'{o}i, Rio de Janeiro, Brazil}

\author{\\Yun-Feng Huang}
\email{hyf@ustc.edu.cn}
\affiliation{CAS Key Laboratory of Quantum Information, University of Science and Technology of China, Hefei, 230026, People’s Republic of China}
\affiliation{Synergetic Innovation Center of Quantum Information and Quantum Physics,University of Science and Technology of China, Hefei, 230026, People’s Republic of China}

\author{Marcelo S. Sarandy}
\email{msarandy@id.uff.br}
\affiliation{Instituto de F\'{i}sica, Universidade Federal Fluminense, Av. Gal. Milton Tavares de Souza s/n, Gragoat\'{a}, 24210-346 Niter\'{o}i, Rio de Janeiro, Brazil}

\author{Chuan-Feng Li}
\email{cfli@ustc.edu.cn}
\affiliation{CAS Key Laboratory of Quantum Information, University of Science and Technology of China, Hefei, 230026, People’s Republic of China}
\affiliation{Synergetic Innovation Center of Quantum Information and Quantum Physics,University of Science and Technology of China, Hefei, 230026, People’s Republic of China}

\author{Guang-Can Guo}
\affiliation{CAS Key Laboratory of Quantum Information, University of Science and Technology of China, Hefei, 230026, People’s Republic of China}
\affiliation{Synergetic Innovation Center of Quantum Information and Quantum Physics,University of Science and Technology of China, Hefei, 230026, People’s Republic of China}

\begin{abstract}
	It is known that high intensity fields are usually required to implement shortcuts to adiabaticity
	via Transitionless Quantum Driving (TQD). Here, we show that this requirement can be relaxed by
	exploiting the gauge freedom of generalized TQD, which is expressed in terms of an arbitrary phase when mimicking the adiabatic evolution.
	We experimentally investigate the performance of generalized TQD in comparison
	with both traditional TQD and adiabatic dynamics. By using a $^{171}$Yb$^+$ trapped ion
	hyperfine qubit, we implement a Landau-Zener adiabatic Hamiltonian and its (traditional and
	generalized) TQD counterparts. We show that the generalized theory provides energy-optimal Hamiltonians for TQD, with no additional fields required. In addition, the
	optimal TQD Hamiltonian for the Landau-Zener model is investigated under dephasing.
	Even using less intense fields, optimal TQD exhibits fidelities that are more
	robust against a decohering environment, with performance superior than that provided by the adiabatic dynamics.
\end{abstract}

\maketitle

\section{Introduction}

Transitionless Quantum Driving (TQD) \cite{Demirplak:03,Demirplak:05,Berry:09} is
a useful technique to mimic adiabatic quantum tasks at finite time.
It has been applied for speeding up adiabaticity in several applications,
such as quantum gate Hamiltonians~\cite{Santos:15,Santos:16,Coulamy:16},
heat engines in quantum thermodynamics~\cite{Adolfo:16},
quantum information processing \cite{Marcela:14,Xia:16,Chen:10}, among
others (e.g., Refs.~\cite{Stefanatos:14,Lu:14,Deffner:16,XiaPRA:14,Song:16,Adolfo:13,Saberi:14,An:16,Zhang:16}).
To perform TQD we need to design a \textit{counter-diabatic} Hamiltonian $H_{\text{CD}}(t)$,
given by $H_{\text{CD}}\left( t\right) =i  \sum_{n}\left( \left\vert
\dot{n}_t\right\rangle \left\langle n_t\right\vert +\left\langle \dot{n}_t|n_t\right\rangle
\left\vert n_t\right\rangle \left\langle n_t\right\vert
\right)$, where $\{|n_t\rangle\}$ denotes the set of instantaneous eigenstates of the original adiabatic
Hamiltonian $H_0(t)$ and the dot symbol denotes time derivative.
The Hamiltonian $H_{\text{CD}}(t)$ enables us to cancel any transition between two different eigenstates
of $H_{0}(t)$. It should be added to the original Hamiltonian $H_0(t)$ to yield the {\it shortcut to adiabaticity}
Hamiltonian, which is given by $H_{\text{SA}}(t) = H_{0}(t)+H_{\text{CD}}(t)$.

In general, the Hamiltonian $H_{\text{SA}}(t)$ allows us to exactly mimic an adiabatic dynamics at arbitrary finite time,
so that, by starting at a given initial eigenstate $|k_0\rangle$ of $H_0(0)$, the evolved state is given by $\ket{ \psi (t)} = e^{i \int_{0}^{t} \theta^{\text{ad}} _{k}\left( \xi \right) d \xi } \ket{k_t}$,
where $\theta^{\text{ad}} (t)= - E_n(t) + i \langle n_t |\dot{n}_t\rangle$ is the adiabatic phase~\cite{Berry:84}.
However, there are several applications where we do not need to exactly mimic an
adiabatic phase $\theta^{\text{ad}} (t)$,
as long as the system is kept in an eigenstate of the Hamiltonian $H_{0}(t)$ \cite{Santos:15,Santos:16,Coulamy:16,Adolfo:16,Marcela:14,Xia:16,Chen:10,Stefanatos:14,Lu:14,Deffner:16,XiaPRA:14,Song:16,Adolfo:13,Saberi:14,An:16,Zhang:16,Liang:16,Yi:17,Ramanathan:16}.
Thus, it has been proposed a generalized approach for TQD, where we consider that the phases that accompany the dynamics can be taken as arbitrary \cite{Torrontegui:13,Santos:18-b}.
Therefore, differently from \textit{traditional} TQD, where the dynamics is driven by $H_{\text{SA}}(t) = H_{0}(t)+H_{\text{CD}}(t)$,
in {\it generalized} TQD the dynamics is driven by a generalized shortcut to adiabaticity Hamiltonian $H_{\text{GSA}}(t)$ written as \cite{Torrontegui:13,Santos:18-b,Chen:11}
\begin{eqnarray}
H_{ \text{\text{GSA}} }\left( t\right) =i \sum\nolimits_{n} \left( \frac{ }{ } |\dot{n}_t\rangle
\langle n_t |+i\theta _{n}\left( t\right) |n_t
\rangle \langle n_t | \frac{ }{ } \right) \text{ , }
\label{HSA}
\end{eqnarray}
where  $\theta_{n}(t)$ are arbitrary real parameters.
In this case, the evolved generalized state is written as $\ket{ \psi (t)} = e^{i \int_{0}^{t} \theta _{k}\left( \xi \right) d \xi } \ket{k_t}$, with arbitrary $\theta _{k}(t)$.
It is worth highlighting that \eqref{HSA} shows that we can mimic an adiabatic task even when our physical system does not allow for a direct implementation of the adiabatic Hamiltonian.
In addition, as it was discussed in \cite{Santos:18-b}, we can also take advantage of these generalized phases $\theta_{n}(t)$ to simplify the Hamiltonian $H_{\text{GSA}}(t)$, possibly even removing in certain situations its dependence on time.

Concerning robustness against decoherence, some experimental and theoretical studies of TQD
in different experimental architectures have shown promising features, such as in nitrogen-vacancy setups~\cite{Liang:16}, trapped ions~\cite{An:16}, atoms in cavities~\cite{Xia:16,Yi:17},
nuclear magnetic resonance (NMR)~\cite{Ramanathan:16}, and optomechanics~\cite{Zhang:16}.
These works consider the traditional approach for shortcuts to adiabaticity, where the adiabatic phase
is taken when mimicking the adiabatic dynamics. As theoretically found out in Ref.~\cite{Santos:18-b},
generalized TQD may provide much better resistance against decoherence depending on the time window designed
to run the quantum process.

In this paper, we report the first experimental implementation aiming at verifying the advantage of generalized
TQD with respect to its adiabatic and traditional TQD counterparts. More specifically, we implement the energetically \textit{optimal} version of generalized TQD in a two-level system realizing the Landau-Zener model.
Our physical system is composed of a single $^{171}$Yb$^+$ ion confined in a Paul trap, where the system
dynamics is driven by the Landau-Zener Hamiltonian.
It has previously been shown that the traditional TQD method requires high intensity fields  \cite{Santos:15,Santos:16,Coulamy:16,Campbell-Deffner:17,Abah:17}.
As we shall see, generalized TQD theory requires much less intense fields while capable of providing better fidelities.
In particular, for the dynamics considered here, while traditional TQD requires additional fields,
the generalized theory allows us to mimic an adiabatic dynamics without additional fields.
Moreover, the dynamics under dephasing shows the superiority of the generalized TQD approach, with higher fidelities with less expenditure of energy resources.

\begin{figure}[t!]
	\centering
	\includegraphics[width=\linewidth]{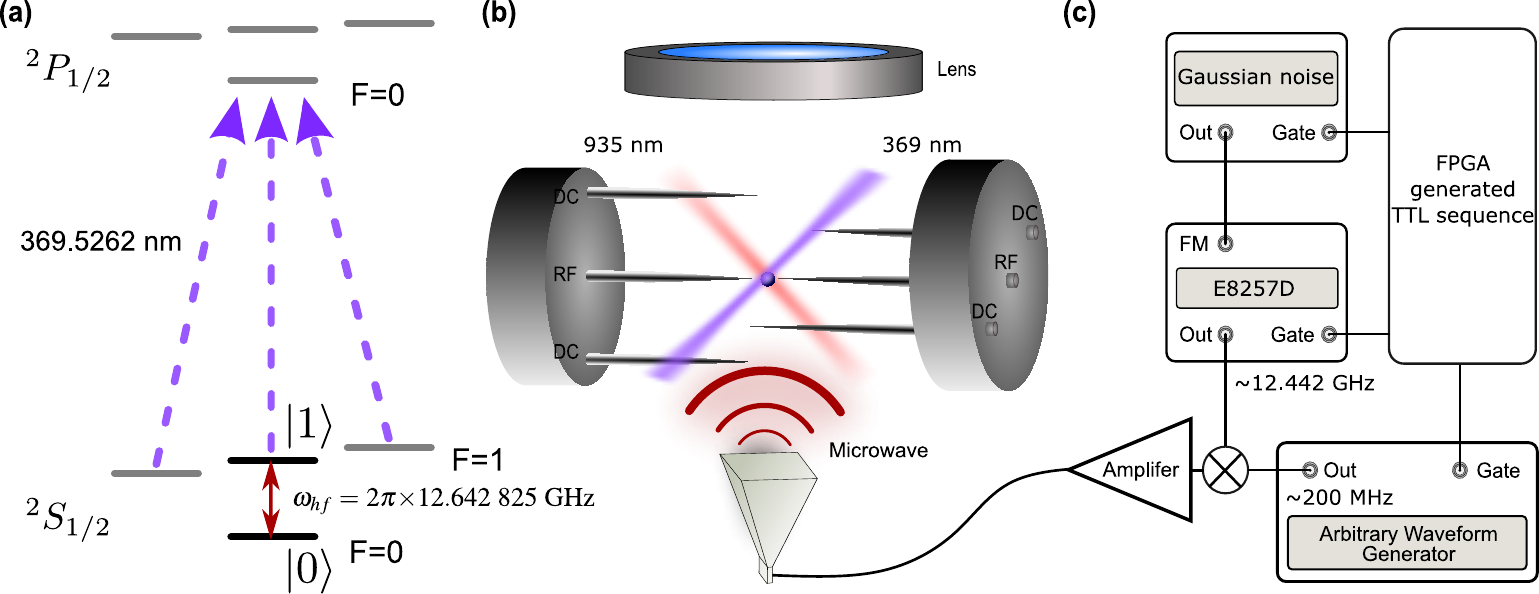}
	\caption{Experimental setup for implementing the generalized TQD. {\color{blue}(a)} The energy spectrum of the $^{171}$Yb$^+$ ion,
		where our two-level system was encoded in the hyperfine energy levels $^{2}S_{1/2}\, \ket{F=0,m_{F}=0}$ and $^{2}S_{1/2}\, \ket{F=1,m_{F}=0}$ and the 369.5 nm laser is used for fluorescence detection.
		{\color{blue}(b)} Diagram of the six needles Paul trap used in our experiment.
		The microwave horn sends out microwaves to drive the hyperfine qubit and the numerical aperture NA = 0.4 objective lens collects the ion fluorescent photon.
		{\color{blue}(c)} Experimental instrument for generating the driving field around $2\pi \times$12.642 GHz. The Gaussian noise source is used to frequency modulate the driving microwave as a dephasing} 
	\label{Fig1}
\end{figure}

\section{Optimal TQD for the Landau-Zener Hamiltonian}

We consider a quantum bit (qubit) undergoing an adiabatic dynamics governed by the
Landau-Zener Hamiltonian, which reads $H_{0}\left( s\right) = - \Delta \sigma _{z} - \Omega_{\text{R}} \left( s\right) \sigma _{x}$, with a detuning $\Delta$ between the microwave frequency and the atom transition level; $\Omega_{\text{R}} \left( s\right)$ is the Rabi frequency and $s = t/\tau$ is the normalized time.
The system is initialized in the ground state $\ket{0}$ of $H_{0}\left( 0\right)$ and adiabatically
evolves to $\ket{\psi(s)} = e^{-i \tau \int_{0}^{s} \theta (\xi) d\xi} \ket{E_{+}(s)}$, where $\ket{E_{+}(s)} = \cos [\vartheta \left( s\right)/2] \ket{0} + \sin [\vartheta \left( s\right)/2] \ket{1}$, with $ \vartheta \left( s\right) = \arctan \left[\Omega_{\text{R}} \left( s\right)/\Delta \right]$ a time-dependent dimensionless parameter that satisfies the boundary conditions $\vartheta \left( 0\right) = 0$ and $\vartheta \left( 1\right) = \vartheta_{0}$. Traditional TQD mimics the above dynamics through the Hamiltonian $
H_{\text{SA}}\left( s\right)$, where $H_{\text{CD}}\left( s\right) =[ d_{s}\vartheta \left( s\right) / 2\tau] \sigma _{y} $, with the function $\vartheta \left( s\right)$ chosen as $\vartheta \left( s\right) = \vartheta_{0} s$ for obtaining a time-independent counter-diabatic Hamiltonian (see, e.g., Ref.~\cite{Santos:18-b}).

However, in this scenario, we still need to implement a time-dependent contribution due to the adiabatic Hamiltonian $H_0(s)$. On the other hand,	
if we adopt generalized TQD, we can implement the transitionless dynamics through the simplest time-dependent Hamiltonian given by $
H_{\text{OpSA}}\left( s\right) = H_{\text{CD}}\left( s\right) $.
As in Ref. \cite{Santos:18-b}, the energetically optimal Hamiltonian above is obtained by choosing the phases in \eqref{HSA} through a geometric contribution, such  as $\theta _{n}\left( t\right) = i \interpro{\dot{n}_{t}}{n_{t}}$,
which implies here in $\theta _{n}\left( t\right) = 0$. The $
H_{\text{OpSA}}\left( s\right)$ turns out to be implemented by a flat $\pi$-pulse, whose robustness has been investigated for fast population transfer in Ref.~\cite{X-Jing:13}.

\section{Experimental setup}

We now demonstrate the experimental implementation of generalized TQD,
with a single $ ^{171}Yb^{+}$ ion trapped in a six needles Paul trap,
which is  shown in Fig. \ref{Fig1}{\color{blue}(b)}.
We encode a qubit into two  hyperfine energy levels of the $^{2}S_{1/2}$ ground state,
which is denoted by $\ket{0} \equiv\, ^{2}S_{1/2}\, \ket{F=0,m_{F}=0}$ and  $\ket{1}\equiv\, ^{2}S_{1/2}\,\ket{F=1,m_{F}=0}$,
as shown in Fig. \ref{Fig1}{\color{blue}(a)}. Applying a 6.40 G static magnetic field,
the clock transition frequency between $\ket{0}$ and $\ket{1}$ is $\omega_{hf} = $ $2\pi \times$12.642 825 GHz.

In a configuration of a frequency mixing scheme with  an arbitrary waveform generator (AWG),
we can coherently drive the hyperfine qubit.
The scheme is shown in  Fig. \ref{Fig1}{\color{blue}(c)}, a 12.442 GHz microwave is mixed with the AWG signals around 200 MHz,
to generate the operating waveform. The AWG waveform is programmed to control $\omega$ and $\Delta$,
for implementing the target Hamiltonian.
In addition, in order to mimic the two level system interacting with an environment,
we introduced a Gaussian  noise  frequency modulation of the $2\pi \times$12.442 GHz microwave, which can be viewed as a dephasing channel.
It is a remarkable fact that the dephasing channel is premium and highly controlled. 
After Doppler cooling of the trapped ion,
we apply a standard optical pumping process to initialize the qubit into the $\ket{0}$ state with 99.9\% efficiency.
After the qubit operation with microwave sequence, a florescence detection method is used to measure the population of the the $\ket{1}$ state \cite{olmschenk_2007}. %
The florescence of the trapped ion is collected by an optical lens with a numerical aperture of NA = 0.4,
then pass an optical bandpass filter and a pinhole, and is finally detected by a PMT with 20\% quantum efficiency.
Within 300 $\mu s$ detection time, the target state preparation and measurement fidelity is measured as 99.4\% \cite{blume-kohout_2017}.

\subsection{Microwave fields and energy resources}

In our experiment we used both resonant and off-resonant microwaves fields to implement adiabatic and TQD dynamics.
The adiabatic dynamics was performed by using a non-resonant microwave,
where the time-dependent effective Rabi frequency is given by $\Omega_{\text{eff}}(s) = [\Omega_{\text{R}}^2(s) + \Delta^2]^{1/2}$. To drive the system by using the traditional TQD,
we use an independent microwave to simulate the counter-diabatic term $H_{\text{CD}}(s)$.
This additional field is a resonant microwave ($\Delta = 0$),
whose Rabi frequency can be obtained as $\Omega_{\text{R-CD}}(s) = d_{s}\vartheta \left( s\right) / 2\tau$.
On the other hand, different from traditional TQD Hamiltonian,
the optimal dynamics driven by $H_{\text{GSA}}(s)$ could be implemented using a single resonant microwave with Rabi frequency $\Omega_{\text{R-CD}}(s)$,
where we have turned-off the non-resonant field used for simulate $H_{0}(s)$.

\begin{figure}[t]
	\centering
	\includegraphics[trim=20 3 20 33,clip,width=0.95\linewidth]{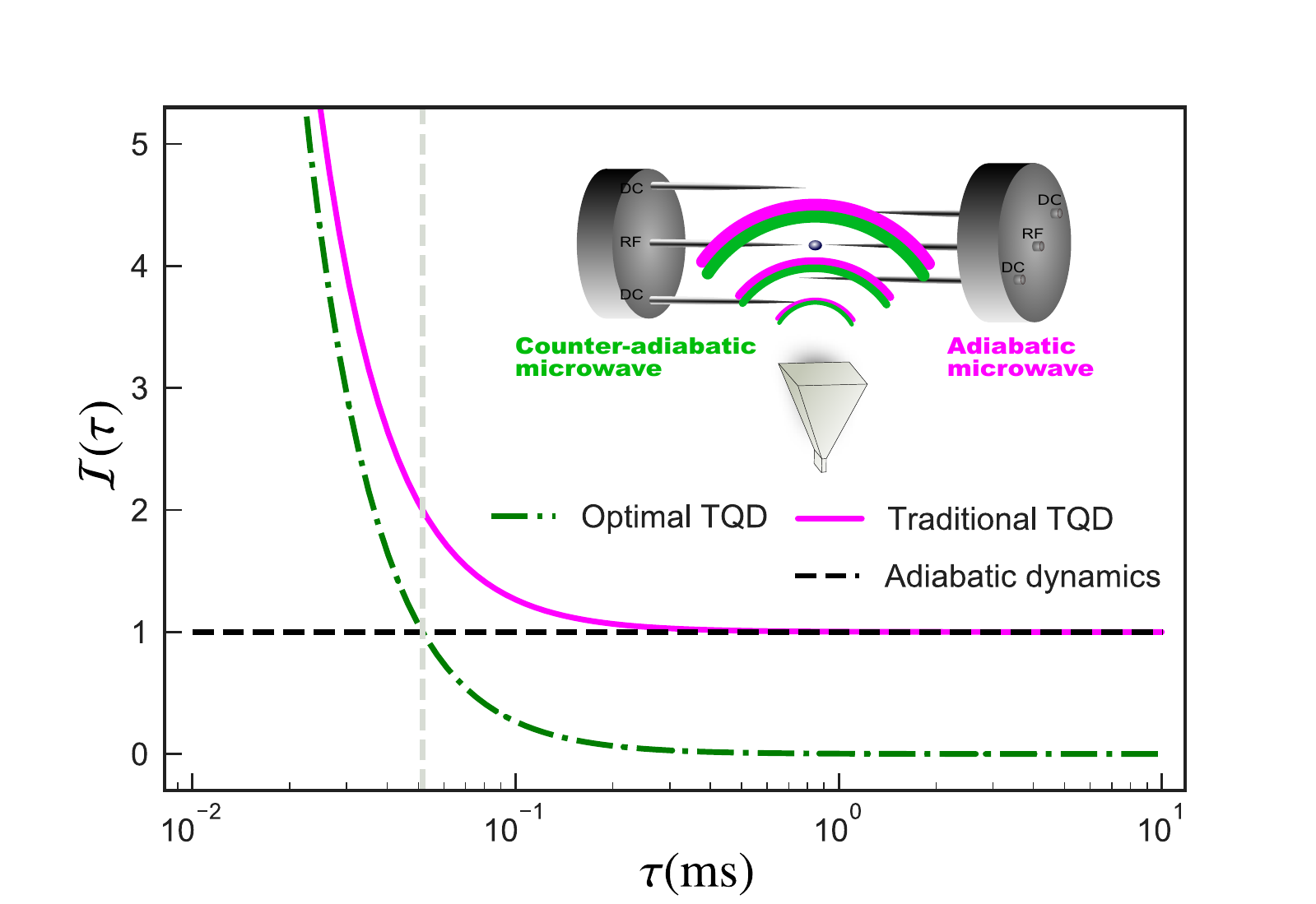}
	\caption{The calculated relative field intensity $\Ical (\tau)$ for traditional TQD $\Ical_{\text{SA}} (\tau)$ (magenta continuum line) and optimal TQD $\Ical_{\text{OpSA}} (\tau)$ (green dashed-dot line) as function of the total evolution time $\tau$, where the horizontal black dashed line represents $\Ical_{\text{Ad}} (\tau)$. The gray vertical line denotes the boundary time $\tau_{\text{B}} \approx 0.052$ ms between the regions $\bar{I}_{\text{OpSA}} (\tau)>\bar{I}_{\text{ad}} (\tau)$ (left hand side) and $\bar{I}_{\text{OpSA}} (\tau)<\bar{I}_{\text{ad}} (\tau)$ (right hand side). Inset: schematic representation of the fields used to implement $H_{0}\left( s\right)$, $H_{\text{SA}}\left( s\right)$ and $H_{\text{OpSA}}\left( s\right)$. We set $\vartheta \left( s\right) = \pi s /3 $, $\Delta = 2\pi \times 2$ KHz and $\Omega_{\text{R}} \left( s\right) = \Delta \tan (\pi s /3)$ is used in our experiment.} \label{Fig2}
\end{figure}

In order to quantify the energy resources employed in the quantum evolution, we study the intensity of the fields used to perform the adiabatic dynamics as well as traditional and optimal TQD.
The field intensity is associated with the Rabi frequency through the relation $I(s) = \Gamma \Omega_{\text{R}}^2(s)$, where $\Gamma$ is a constant that depends on the microwave amplifier.
Considering the whole evolution time, we can define the average intensity field as $\bar{I} (\tau) = (1/\tau) \int \nolimits _{0}^{\tau} I(t) dt = \int \nolimits _{0}^{1}  I(s) ds$.
So, for the adiabatic dynamics we have $\bar{I}_{\text{Ad}} (\tau) = \Gamma \int \nolimits _{0}^{1} \Omega_{\text{R}}^2(s) ds$, for the optimal TQD we get $\bar{I}_{\text{OpSA}} (\tau) = \Gamma \int \nolimits _{0}^{1} \left[d_{s}\vartheta \left( s\right) / 2\tau \right]^2 ds$, and for the traditional TQD we find $\bar{I}_{\text{SA}} (\tau) = \bar{I}_{\text{Ad}} (\tau) + \bar{I}_{\text{OpSA}} (\tau)$, since
the traditional TQD field is composed by both the adiabatic and the optimal TQD contributions.
For convenience, we disregard the constant $\Gamma$ by taking relative field intensities expressed in unities of the adiabatic intensity $\bar{I}_{\text{Ad}}$.
Then, we define $\Ical_{\text{SA}} (\tau) = \bar{I}_{\text{SA}} (\tau) / \bar{I}_{\text{Ad}}$ and $\Ical_{\text{OpSA}} (\tau) = \bar{I}_{\text{OpSA}} (\tau) / \bar{I}_{\text{Ad}}$, and adopt the normalization $\Ical_{\text{Ad}} (\tau) = 1$. These field intensities are plotted in Fig~\ref{Fig2}{\color{blue}}, with an schematic representation of each dynamics indicated in the
inset. As we can see from Fig. \ref{Fig2}, we can define a value $\tau_{\text{B}}$ for the total evolution time for which the intensity fields for implementing optimal TQD becomes less intense than the adiabatic intensity. By computing $\tau_{\text{B}}$ we get $\tau_{\text{B}} \approx 0.052$ ms and we represent this boundary using a vertical line in Fig.~\ref{Fig2}. Notice that, after $\tau_{\text{B}}$, the shortcut to adiabaticity defined by the optimal TQD can be
implemented by spending less energy resources, as measured by the field intensity, than the adiabatic approach.

An alternative approach to compute the energy cost is to estimate the energy scale through the
time-average Hamiltonian norm~\cite{Santos:15,Santos:16,Santos:18-b,Coulamy:16,Nathan:14}. In this direction,
let us consider a quantum dynamics driven by a time-dependent Hamiltonian, with energy cost defined as
$\Sigma (\tau) = \int \nolimits _{0}^{\tau} \sqrt{\text{Tr} \{ H^2(t) \} } dt$.
For the specific case of the Landau-Zener model, this method takes into account not only the
external field intensities but also the detuning $\Delta$. Indeed, $\Delta$ exerts influence over
the energy gap spectrum, which may justify its accounting in some scenarios, such as adiabatic quantum  computation.
From this definition, we can show that $\Sigma_{\text{SA}} (\tau) \geq \Sigma_{\text{Ad}} (\tau)$, for every $\tau$.
Moreover, there is again a boundary value $\tau_{\text{B,}\Sigma}$ for which we get $\Sigma_{\text{Op-SA}} (\tau) < \Sigma_{\text{Ad}} (\tau)$ for $\tau > \tau _{\text{B,}\Sigma}$,
as originally predicted  in Ref.~\cite{Santos:18-b}.
Here, we obtain $\tau _{\text{B,}\Sigma} \approx 0.033$ ms.
Thus, we can see that an analysis from energy scale of the Hamiltonian provides a (qualitatively) equivalent result to the intensity ields analysis.
In both cases, it is shown that,
by specifically designing a suitable time window, shortcuts to adiabaticity can be implemented in such a way to accelerate physical processes while saving energy resources.

\begin{figure}[t!]
	\centering
	\includegraphics[trim=10 3 29 33,clip,width=0.95\linewidth]{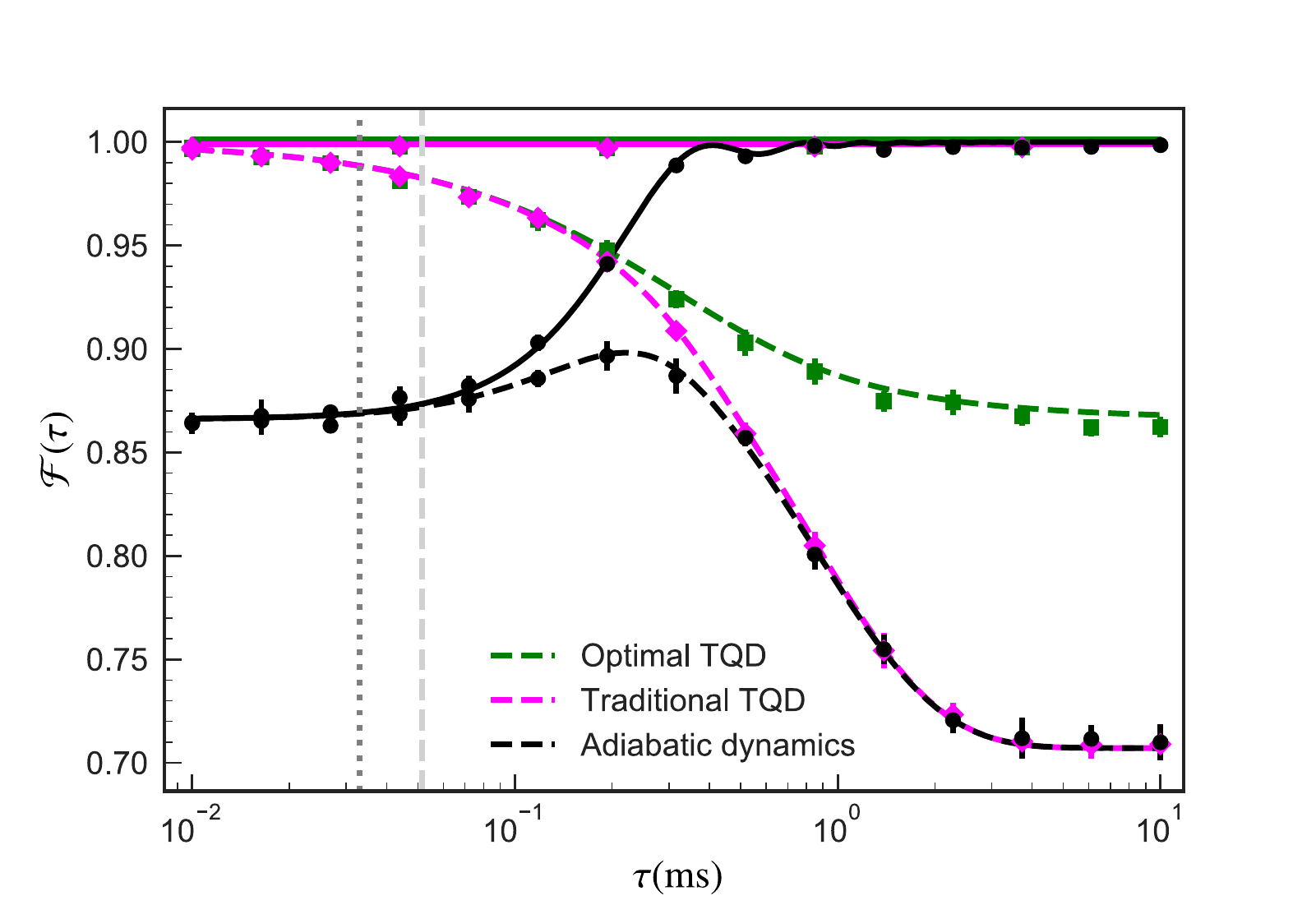}
	\caption{Fidelity for unitary dynamics (continuum lines) and non-unitary one (dashed lines) under dephasing. The symbols and lines represent experimental data and theoretical results, respectively. We set the Hamiltonians parameter as in Fig. \ref{Fig2}
		, while the decohering rate was kept as $\gamma = 2.5$ KHz for all non-unitary dynamics. The first gray vertical line denotes the boundary time $\tau_{\text{B}} \approx 0.052$ ms as discussed in Fig. \ref{Fig1}, while the second vertical lines is associated with $\tau _{\text{B,}\Sigma} \approx 0.033$ ms.} \label{Fig3}
\end{figure}

\section{Robustness against decoherence}
Owing to the long coherence time of the $^{171}$Yb$^+$ trapped ion hyperfine qubit, the decoherence effects for the timescale adopted in the experimental setup is negligible.
Therefore, to simulate the interaction of the Landau-Zener system with an environment, we introduce a Gaussian noise capable of implementing a dephasing channel where the system will be driven by a Lindblad equation given by
$\dot{ \rho }\left( t\right) =-i \left[ H \left(
t\right) ,\rho \left( t\right) \right] + \gamma \left[ \sigma_{z}\rho
\left( t\right)\sigma_{z} - \rho
\left( t\right) \right]$
, where $\gamma$ is the dephasing rate.
To quantify the robustness of the protocols,
	we use the fidelity $\Fcal (\tau) = \sqrt{\bra{\psi(s)}\rho\left( s\right) \ket{\psi(s)}}$, which is related to the Bures length \cite{Bures:69}, with $\rho\left( s\right)$ being solution of Lindblad equation and $\ket{\psi(s)}=\ket{E_{+}(s)}$.

We start by disregarding decoherence and implementing each protocol for several choices of $\tau \in [10^{-2}\,$ms$,10\,$ms$]$. The results are shown in Fig. \ref{Fig3}. Since the maximum total evolution time is $\tau=10$ ms (i.e., $20$ times shorter than the coherence time of the qubit), the dynamics can be considered as approximately unitary. The continuum lines in Fig. \ref{Fig3} show the fidelities $\Fcal (\tau)$ under a unitary dynamics for each protocol and the corresponding experimental data points. The black curve shows that the adiabatic behavior is achieved for long total evolution time $\tau \gg \tau_{\text{ad}} \approx 0.021$\,ms, with $\tau_{\text{ad}}$ computed from $\tau_{\text{ad}} = \max_{s\in [0,1]}|\bra{E_{-}(s)} d_{s} H_{0}(s)\ket{E_{+}(s)}/g^{2}(s)|$, where $g(s) = E_{+}(s)-E_{-}(s)$ is the gap between fundamental $\ket{E_{-}(s)}$ and excited $\ket{E_{+}(s)}$ energy levels \cite{Sarandy:04}.
Notice that, for fast evolutions ($\tau < 0.021\,$ms), both optimal and traditional TQD provide a high fidelity protocol, while adiabatic dynamics fails. Naturally, this high performance is accompanied by costly fields used for implementing the TQD protocols, as previously shown in Fig. \ref{Fig2}{\color{blue}}. On the other hand, it is important to highlight that for total evolution time $\tau > \tau_{\text{B}}$, where the optimal TQD field is smaller than both adiabatic and traditional TQD ones, the high performance of optimal TQD is kept. By looking now at the effect of the dephasing channel, we can see that optimal TQD exhibits the highest robustness for every $\tau$, with its advantage increasing as $\tau$ increases.
For fast evolution times, this high performance is again associated with costly fields in comparison with the
adiabatic fields.
However, for large evolution times, while the traditional TQD fidelity converges to adiabatic fidelity, the fidelity behavior of optimal TQD is much better than both adiabatic dynamics
and traditional TQD, with less energy resources spent in the process. This is a remarkable result,
since we can achieve both better fidelities with less energy fields involved, 
where we get $\Fcal_{\text{OpSA}} > \Fcal_{\text{Ad}}$ even when $\bar{I}_{\text{OpSA}} \ll \bar{I}_{\text{Ad}}$.

\section{Conclusion}

In this paper we have experimentally investigated the performance of shortcuts
to adiabaticity by exploiting the gauge freedom as we fix the phase accompanying the evolution dynamics.
In particular, we have focused on the optimal TQD, comparing it with its associated adiabatic dynamics and traditional TQD counterparts.
Our main results are: i) The optimal version of generalized TQD has been shown to be a useful protocol for obtaining the optimal shortcut to adiabaticity.
While adiabatic and traditional TQD require time-dependent quantum control, optimal TQD can be experimentally realized by using time-independent fields.
In addition, the necessity of auxiliary fields in traditional TQD is not a requirement for implementing TQD via its optimal version.
ii) Optimal TQD is an energetically optimal protocol of shortcut to adiabaticity.
By considering the average intensity fields as a measure of energy cost for implementing
the protocols discussed here, we were able to show that the optimal version of generalized TQD may be energetically less demanding.
This result is kept also for alternatives definitions of energy cost, e.g.
taking into account the detuning contribution.
iii) By simulating an environment-system coupling associated with the dephasing channel,
we have shown that optimal TQD can be more robust than the adiabatic dynamics and the traditional TQD while
at the same time spending less energy resources for a finite time range. In addition, we were able to to mimic the adiabatic behavior through time-independent Hamiltonian, showing that the optimal theory works.
Once the Landau-Zener Hamiltonian can be implemented in others physical systems, e.g. nuclear magnetic resonance \cite{Nielsen:Book} and two-level systems driven by a chirped field \cite{Demirplak:03},
it is reasonable to think that the results obtained here can be also realized in other experimental architectures.
Moreover, inverse engineering protocols are an interesting candidate to implement fast and robust elementary quantum gates for quantum computing \cite{Santos:15,Santos:16,Coulamy:16,Santos:18-a}. Thus, an experimental investigation of these protocols can be a promising direction as a future research, as can the robustness of optimal TQD against others classes of errors (as it has been done for traditional TQD in Ref.~\cite{Ruschhaupt:12}).


\section{Funding Information}

This work was supported by the National Key Research and Development Program of China (No. 2017YFA0304100), National Natural Science Foundation of China (Nos. 61327901, 61490711,11774335, 11734015, 11474268, 11374288, 11304305,11404319), Anhui Initiative in Quantum Information Technologies (AHY070000), Key Research Program of Frontier Sciences, CAS (No. QYZDY-SSWSLH003), the National Program for Support of Top-notch Young Professionals (Grant No. BB2470000005), the Fundamental Research Funds for the Central Universities (WK2470000026). A.C.S. is supported by CNPq-Brazil. M.S.S. is supported by CNPq-Brazil (No. 303070/2016-1), FAPERJ (No. 203036/2016), and the Brazilian National Institute
for Science and Technology of Quantum Information (INCT-IQ).

\bibliography{/home/alan/Dropbox/School/Articles/Gaveta/Models/Bibliografia/mybib-noURL.bib}
\bibliographystyle{Phys-Rev-A}

\end{document}